# The Phonon Quasiparticle Approach for Anharmonic Properties of Solids

Zhen Zhang[1], Dong-Bo Zhang[2,3], Tao Sun[4], and Renata M. Wentzcovitch[1,5,6,*]

[1]Department of Applied Physics and Applied Mathematics, Columbia University, New York, NY 10027, USA
[2]College of Nuclear Science and Technology, Beijing Normal University, Beijing 100875, People's Republic of China
[3]Beijing Computational Science Research Center, Beijing 100193, People's Republic of China
[4]Key Laboratory of Computational Geodynamics, Chinese Academy of Sciences, Beijing 100049, People's Republic of China
[5]Department of Earth and Environmental Sciences, Columbia University, New York, NY 10027, USA
[6]Lamont–Doherty Earth Observatory, Columbia University, Palisades, NY 10964, USA

[*]rmw2150@columbia.edu

**Abstract**. Knowledge of lattice anharmonicity is essential to elucidate distinctive thermal properties in crystalline solids. Yet, accurate *ab initio* investigations of lattice anharmonicity encounter difficulties owing to the cumbersome computations. Here we introduce the phonon quasiparticle approach and review its application to various materials. This method efficiently and reliably addresses lattice anharmonicity by combining *ab initio* molecular dynamics and lattice dynamics calculations. Thus, in principle, it accounts for full anharmonic effects and overcomes finite-size effects typical of *ab initio* molecular dynamics. The validity and effectiveness of the current approach are demonstrated in the computation of thermodynamic and heat transport properties of weakly and strongly anharmonic systems.

## 1. Introduction

In crystalline materials, anharmonicity describes the deviation of vibrations from harmonic oscillations and is caused by phonon-phonon interactions. Lattice anharmonicity is essential in understanding phonon-driven phenomena, e.g., structural phase transitions [1-5], thermal expansion [6-8], and lattice thermal conductivity [9]. Lattice anharmonicity can also be characterized by the intrinsic temperature dependence of phonon frequencies. In terms of *ab initio* calculations, there are two prevailing ways to deal with lattice anharmonicity. The first is the interatomic force-constant-based methods, which can evaluate anharmonic phonon properties and material's thermal properties with dense **q**-meshes including essential long-wavelength phonons. However, the higher-order (3$^{rd}$, 4$^{th}$, etc.) interatomic force constants are complicated and computationally heavy to obtain. Inevitably, it suffers from computational errors or truncation issues. The second is the molecular-dynamics-based methods. Free energies and thermodynamic properties can be obtained by thermodynamic integration (TI). However, to approach the thermodynamic limit ($N \to \infty$), conducting TI using *ab initio* molecular dynamics (AIMD) simulations with a sufficiently large supercell is beyond the current computational capability.

The present phonon quasiparticle approach deals with lattice anharmonicity by extracting phonon quasiparticle properties, i.e., renormalized frequencies and phonon lifetimes, from AIMD simulations without explicitly expressing the higher-order interatomic force constants. This is done by computing the normal-mode-projected velocity autocorrelation function (VAF) obtained by AIMD. Renormalized frequencies are next Fourier interpolated over the whole Brillouin zone (BZ) to overcome finite-size effects. Then thermodynamic properties in the thermodynamic limit ($N \to \infty$) can be obtained within the framework of the phonon gas model (PGM). Since the phonon quasiparticle approach is based on AIMD, in principle, anharmonicity is accounted for to all orders in perturbation theory. This approach can be used to study various thermal properties, e.g., structural stability, thermodynamic properties, phase boundaries, and lattice thermal conductivities.

## 2. Method

We define a phonon quasiparticle numerically by the normal-mode-projected velocity autocorrelation function (VAF) [4, 10],

$$\langle V_{\mathbf{q}s}(0) \cdot V_{\mathbf{q}s}(t) \rangle = \lim_{\tau \to \infty} \frac{1}{\tau} \int_0^\tau V_{\mathbf{q}s}^*(t') V_{\mathbf{q}s}(t' + t) dt', \quad (1)$$

where $V_{\mathbf{q}s}(t) = \sum_{i=1}^{N} \sqrt{M_i} \mathbf{v}_i(t) e^{i\mathbf{q}\cdot\mathbf{R}_i} \cdot \hat{\mathbf{e}}_{\mathbf{q}s}$ is the mass-weighted mode-projected velocity for normal mode ($\mathbf{q}$, s). $\mathbf{q}$ is the phonon wave vector, and *s* indexes the *3n* phonon branches of an *n*-atom primitive cell. $M_i$, $\mathbf{R}_i$, and $\mathbf{v}_i$ ($i = 1, \ldots, N$) are the atomic mass, the atomic equilibrium coordinate, and the atomic velocity computed by AIMD simulations of an *N*-atom supercell, respectively. $\hat{\mathbf{e}}_{\mathbf{q}s}$ is the harmonic phonon polarization vector determined by the harmonic phonon calculations. $\mathbf{q}$ is commensurate with the supercell size. For a well-defined phonon quasiparticle, the VAF can be phenomenologically described as an exponentially decaying cosine function,

$$\langle V_{\mathbf{q}s}(0) \cdot V_{\mathbf{q}s}(t) \rangle = A_{\mathbf{q}s} \cos(\widetilde{\omega}_{\mathbf{q}s} t) e^{-\Gamma_{\mathbf{q}s} t}, \quad (2)$$

where $A_{\mathbf{q}s}$ is the oscillation amplitude, $\widetilde{\omega}_{\mathbf{q}s}$ is the renormalized phonon frequency, and $\Gamma_{\mathbf{q}s}$ is the phonon linewidth inversely proportional to the lifetime, $\tau_{\mathbf{q}s} = 1/(2\Gamma_{\mathbf{q}s})$ [10, 11]. The corresponding power spectrum,

$$G_{\mathbf{q}s}(\omega) = \left| \int_0^\infty \langle V_{\mathbf{q}s}(0) \cdot V_{\mathbf{q}s}(t) \rangle e^{i\omega t} dt \right|^2 \quad (3)$$

should have a Lorentzian line shape with a single peak at $\widetilde{\omega}_{\mathbf{q}s}$ and a linewidth of $\Gamma_{\mathbf{q}s}$ [10, 11]. $\widetilde{\omega}_{\mathbf{q}s}$ can be used to calculate the anharmonic thermodynamic properties. To overcome the finite-size effects on the vibrational contribution to the free energy and to obtain the free energy in the thermodynamic limit, the anharmonic phonon spectrum is further obtained via Fourier interpolation. At each $\mathbf{q}$, we construct the effective harmonic dynamical matrix [4, 10],

$$\widetilde{D}(\mathbf{q}) = [\hat{\mathbf{e}}_{\mathbf{q}}] \Omega_{\mathbf{q}} [\hat{\mathbf{e}}_{\mathbf{q}}]^\dagger, \quad (4)$$

where the diagonal matrix $\Omega_{\mathbf{q}} = \text{diag}[\widetilde{\omega}_{\mathbf{q}1}^2, \widetilde{\omega}_{\mathbf{q}2}^2, \ldots, \widetilde{\omega}_{\mathbf{q}3n}^2]$ contains $\widetilde{\omega}_{\mathbf{q}s}^2$ in the diagonal, and the matrix $[\hat{\mathbf{e}}_{\mathbf{q}}] = [\hat{\mathbf{e}}_{\mathbf{q}1}, \hat{\mathbf{e}}_{\mathbf{q}2}, \ldots, \hat{\mathbf{e}}_{\mathbf{q}3n}]$ contains harmonic polarization vectors in columns. Next, the effective harmonic force constant matrix is obtained by Fourier transform,

$$\widetilde{\Phi}(\mathbf{r}) = \sum_{\mathbf{q}} \widetilde{D}(\mathbf{q}) \cdot e^{i\mathbf{q}\cdot\mathbf{r}}, \quad (5)$$

where anharmonic interactions are effectively captured. Then, $\widetilde{\omega}_{\mathbf{q}'s}$ at arbitrary wave vector $\mathbf{q}'$ in the Brillouin zone (BZ) can be obtained by diagonalizing

$$\widetilde{D}(\mathbf{q}') = \sum_{\mathbf{r}} \widetilde{\Phi}(\mathbf{r}) \cdot e^{-i\mathbf{q}'\cdot\mathbf{r}}, \quad (6)$$

from which the *T*-dependent phonon quasiparticle spectrum can be computed. Within the PGM, the vibrational entropy formula with *T*-dependent phonon quasiparticle spectra is still applicable [4, 10, 11],

$$S_{\text{vib}}(T) = k_B \sum_{\mathbf{q}s} [(n_{\mathbf{q}s} + 1) \ln(n_{\mathbf{q}s} + 1) - n_{\mathbf{q}s} \ln n_{\mathbf{q}s}], \quad (7)$$

where $n_{\mathbf{q}s} = [\exp(\hbar \widetilde{\omega}_{\mathbf{q}s}(T)/k_B T) - 1]^{-1}$. Then the Helmholtz free energy $F(T)$ at any $T$ can be obtained by definition [12, 13],

$$F(T) = U(T) - TS_{\text{el}}(T) - TS_{\text{vib}}(T), \quad (8)$$

where $U(T)$ is the ensemble average of internal energy, i.e., potential (total) energy plus kinetic energy, of the system at finite $T$ obtained from AIMD. $S_{el}$ is the electronic entropy for a metallic system obtained using the Mermin functional [14]. Alternatively, with $F(T_0)$ at a reference temperature $T_0$ is known, $F$ at any other $T$ can also be obtained by integrating the entropy [5, 15],

$$F(T) = F(T_0) - \int_{T_0}^{T} S_{el}(T')dT' - \int_{T_0}^{T} S_{vib}(T')dT'. \quad (9)$$

The reference $F(T_0)$ can be obtained by equation (8) at $T_0$. Both approaches (equations (8) and (9)) have been tested to be effective [5, 12, 13, 15]. In contrast to $S_{vib}$, $S_{el}$ and $U$ depend only weakly on simulation cell size [13]. Therefore, $S_{el}$ and $U$ can be readily obtained from the direct output of AIMD calculations in finite-size cells. Apart from $\widetilde{\omega}_{\mathbf{q}s}$, $\tau_{\mathbf{q}s}$ can be used to calculate the lattice thermal conductivity within the relaxation-time approximation (RTA) of the linearized Boltzmann transport equation (LBTE) [16-18],

$$\kappa_{lat} = \frac{1}{3} \sum_{\mathbf{q}s} c_{\mathbf{q}s} v_{\mathbf{q}s}^2 \tau_{\mathbf{q}s}, \quad (10)$$

where $c_{\mathbf{q}s}$ is the mode heat capacity, and $v_{\mathbf{q}s}$ is the mode group velocity. The latter is obtained by computing $\mathbf{q}$-gradients of $T$-dependent phonon quasiparticle dispersions. The details of the phonon quasiparticle approach and code to perform phonon quasiparticle calculations can be found in [19].

## 3. Applications

### 3.1. Thermodynamics

The method has been successfully applied to weakly [10] and strongly [4, 5, 13, 15] anharmonic systems, nonmetallic [4, 10, 15, 19] and metallic [5, 12, 13] anharmonic systems, and used to study structural stability [4, 5, 20-22], anharmonic phonon dispersions [4, 5, 10, 12, 13, 19], anharmonic free energies [4, 10], phase transitions [5, 12, 13], and anharmonic thermodynamic quantities [15].

*3.1.1. Weakly anharmonic insulating MgSiO$_3$ perovskite.* Pbnm MgSiO$_3$ perovskite (MgPv) is the most abundant phase of the Earth's mantle. Despite the successful application of quasiharmonic approximation (QHA) to such weakly anharmonic systems, intrinsic anharmonic effects have been identified in the Raman spectrum of MgPv at ambient pressure ($P$) [23-25]. By conducting phonon quasiparticle calculations, we obtained frequency shifts $\Delta\omega_{\mathbf{q}s} = \widetilde{\omega}_{\mathbf{q}s} - \omega_{\mathbf{q}s}$ at constant volume ($V$) for MgPv, where $\omega_{\mathbf{q}s}$ is the harmonic phonon frequency, as shown in figures 1(a)–(c). Then $\Delta\omega_{\mathbf{q}s}$ at constant volume were converted into $\Delta\omega_{\mathbf{q}s}$ at constant $P$ by using the quasiharmonic thermal expansivity and mode Grüneisen parameter, which are shown in figures 1(d)–(f). We successfully reproduced the irregular temperature induced frequency shifts [23-25] observed in the Raman spectrum of this complex system [10].

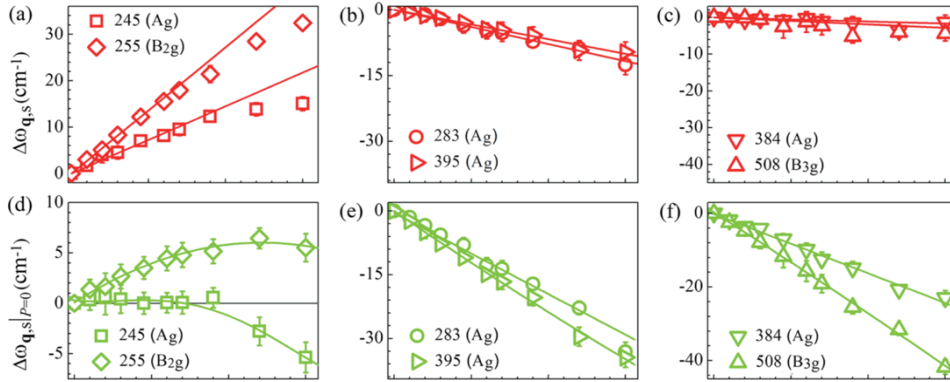

**Figure 1.** (a)–(c) show the $T$-dependent frequency shifts at constant volume $\Delta\omega_{\mathbf{q}s}|_{V(P=0)}$ of MgPv modes with positive, negative, and nearly zero shifts, respectively. (d)–(f) show frequency shifts at zero pressure $\Delta\omega_{\mathbf{q}s}|_{P=0}$. (Results used with permission from [10], Copyright APS.)

As long as phonon quasiparticles exist, i.e., as long as anharmonicity does not mix harmonic modes, and the power spectra of all modes have well-defined single peaks, one can also obtain anharmonic dispersions over the entire BZ, as shown in figure 2(a). The anharmonic phonon dispersion at 300 K was also compared with the harmonic one obtained by density-functional perturbation theory (DFPT) [26]. The frequency shifts are not significant, in line with the fact that MgPv is weakly anharmonic. Their corresponding vibrational densities of states (VDoS) are shown in figure 2(b). Combining these thoroughly sampled quasiparticle dispersions with the phonon gas model can offer first-principles free energy and thermodynamics properties in the thermodynamic limit at temperatures beyond the limit of validity of the QHA [10]. Figure 2(c) shows that free energy obtained from the phonon quasiparticle approach plus PGM agrees well with that obtained from thermodynamic integration (TI) with the same coarse **q**-mesh, which validates the present methodology. Fourier interpolation also allowed free energy computation with a sufficiently dense **q**-mesh, which is discernible from the one obtained with the coarse **q**-mesh. Therefore, the present method overcomes finite-size effects on computations of thermodynamic properties using AIMD simulations. This is especially important for calculations of thermal expansivity and phase boundaries, which are fundamental properties in geodynamics simulations [10].

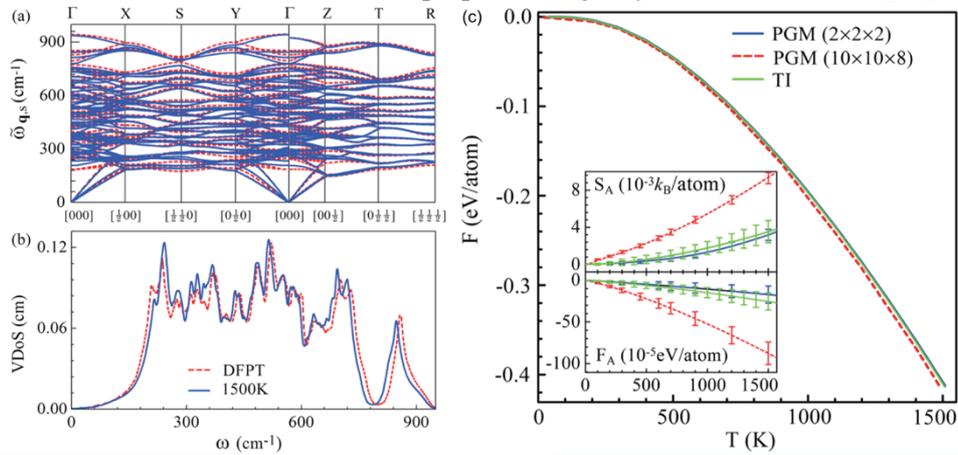

**Figure 2.** (a) Anharmonic phonon dispersions (solid blue lines) of MgPv obtained at $T$ = 1500 K. Harmonic phonon dispersions calculated using DFPT at the same volume (red dashed lines) are shown for comparison. (b) $T$-dependent VDoS of MgPv obtained on a 10×10×8 **q**-mesh using DFPT (red dashed line) and at 1500 K (solid blue line), both at the same volume. The thermal pressure is ~10 GPa. (c) Vibrational free energy $F$ and (insert) excess anharmonic vibrational free energy ($F_A$) and entropy ($S_A$) obtained with the PGM and the TI approach using an MD run sampling a coarse **q**-mesh (2×2×2). Results with a finely interpolated **q**-mesh (10×10×8) use the PGM. Error bars are evaluated from 5 parallel replicas (60 ps each) of Born-Oppenheimer molecular dynamics (BOMD) [27]. (Results used with permission from [10], Copyright APS.)

*3.1.2. Strongly anharmonic insulating CaSiO₃ perovskite.* CaSiO$_3$ perovskite (CaPv) is the third most abundant mineral of the Earth's lower mantle (LM) (~7 vol %). It is also the last major lower mantle phase whose structural and elastic properties have not been well characterized. It is widely believed that under the LM conditions, 23 < $P$ < 135 GPa and 2000 < $T$ < 4000 K, the cubic phase with $Pm\bar{3}m$ space group is adopted by CaPv. However, static calculations for cubic CaPv yield double-well potential and harmonic phonon calculations yield imaginary phonon frequencies around **R** (1/2, 1/2, 1/2) and **M** (0, 1/2, 1/2). At low temperatures, e.g., $T$ < 500 K, CaPv adopts a variety of tetragonal or orthorhombic phases. We used the phonon quasiparticle approach to study cubic CaPv's structural stability and the corresponding phonon spectra. AIMD simulations were carried out for over 50 ps, and the local-density approximation (LDA) was chosen for the exchange-correlation functional. Figure 3(a) shows oxygen atom displacements distributions from AIMD at various $T$. Figure 3(b) shows the phonon spectra for the normal mode with imaginary harmonic frequency at **R** at the same $T$ as in figure 3(a). At $T$ < 600 K, the

O exhibits double-peak distributions with respect to the cubic equilibrium position, meaning the cubic phase is distorted. The corresponding power spectra display double peaks, indicating the breakdown of the phonon quasiparticle. At $T = 600$ K and above, the O shows centered distribution at the cubic equilibrium position, meaning the cubic phase is dynamically stabilized by anharmonic interactions. The corresponding phonon spectrum is well-defined with a single peak [4].

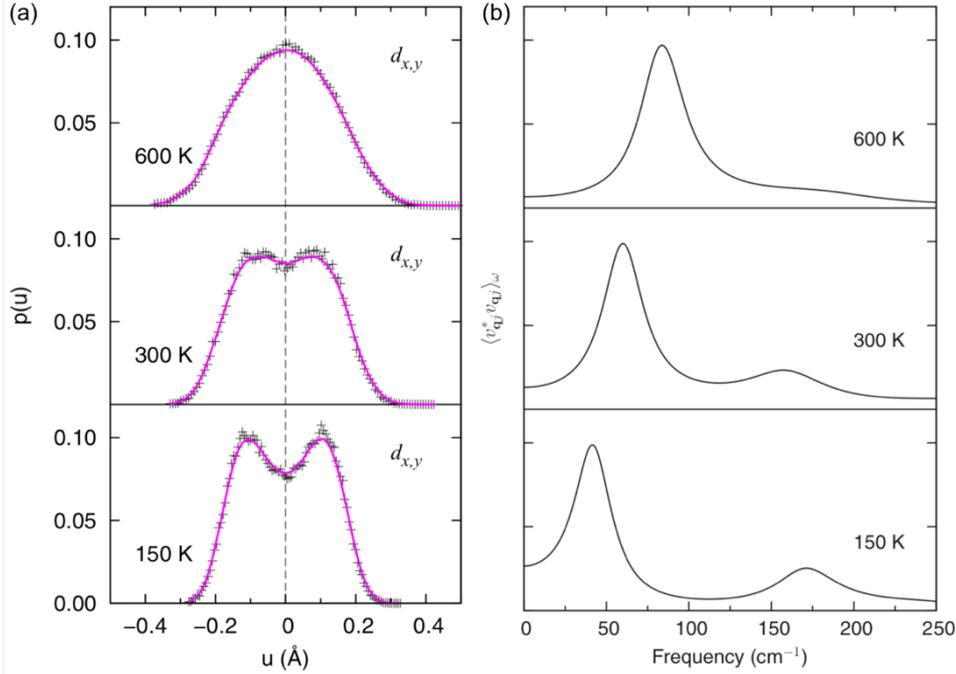

**Figure 3.** (a) Distributions of the O displacements in the $x$ or $y$ direction (octahedral tilting) with respect to the ideal cubic structure ($a^0a^0a^0$) at various $T$. (b) Power spectra of the VAF of $R_{25}$ with **q** vector **R** (1/2, 1/2, 1/2) at low $T$. (Results used with permission from [4], Copyright APS.)

Using a 2×2×2 supercell, the anharmonic entropy calculated using the present phonon quasiparticle spectrum plus the PGM approach and the one by TI are in good agreement [4]. By Fourier interpolation, the high-$T$ anharmonic phonon dispersion for cubic CaPv is free of imaginary frequencies, as shown in figure 4(a), which enables us to compute the vibrational entropy and free energy in the thermodynamic limit. The anharmonic phonon dispersion obtained by the present method also compares well with that obtained by the effective harmonic potential method [3], which further validates the present approach. The advantage of the present method is that complete information on the dynamics of the phonon quasiparticle is given [4]. At high $T$, where the cubic phase is adopted, we carried out phonon quasiparticle calculations for a series of $V$ and $T$. Vibrational entropy and Helmholtz free energy were obtained via equations (7) and (9) and are shown in figures 4(b) and 4(c), respectively [15].

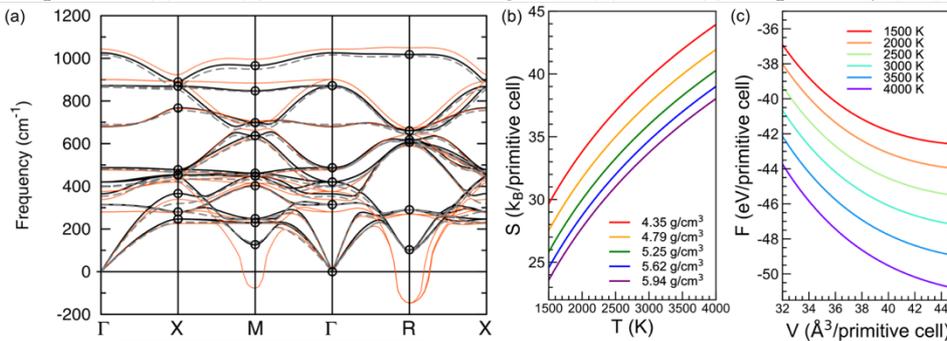

**Figure 4.** (a) Phonon dispersions at 1000K. The circles represent the frequencies extracted from the normal-mode-projected VAF. Fourier interpolating these frequencies yields the renormalized phonon

dispersions, shown in solid black curves. Dashed curves are the phonon dispersions determined by fitting an effective harmonic force constant matrix [3]. Harmonic phonon dispersions calculated with DFPT are shown in orange (light) curves. (Results used with permission from [4], Copyright APS.) (b) Vibrational entropy and (c) Helmholtz free energy of cubic CaPv. (Results used with permission from [15], Copyright APS.)

With Helmholtz free energy obtained, various thermodynamic properties such as thermal expansivity, thermodynamic Grüneisen parameter, isobaric heat capacity, and adiabatic bulk modulus were computed and are shown in figure 5. Results were also compared with a previous work using the Mie-Grüneisen-Debye (MGD) formulation [28], which may not account for full anharmonicity. The differences between the two methods are discernible for all thermodynamic quantities, suggesting the importance of full anharmonic effects. The difference is significant for the Grüneisen parameter, which can be used as a straightforward indicator of anharmonic effects [15]. Note the MGD formulation is based on the fitting for the Mie-Grüneisen equation of state. Its obtained Grüneisen parameter ($\gamma_{mg}$) and the thermodynamic Grüneisen parameter ($\gamma$) of this study are not strictly equal. The two quantities coincide with each other [29, 30] when satisfying three criteria [15]. First, the system is within the framework of QHA. Second, all mode Grüneisen parameters ($\gamma_i$) are equal. Third, at sufficiently high temperature, all mode isochoric heat capacities ($C_{Vi}$) are equal, and all mode thermal energies ($E_{thi}$) are equal. Here for strongly anharmonic cubic CaPv at finite temperatures, none of the criteria is satisfied, resulting in a difference between $\gamma$ and $\gamma_{mg}$. A detailed discussion on the difference and the relation among the three criteria can be found in the last paragraph of section III of [15].

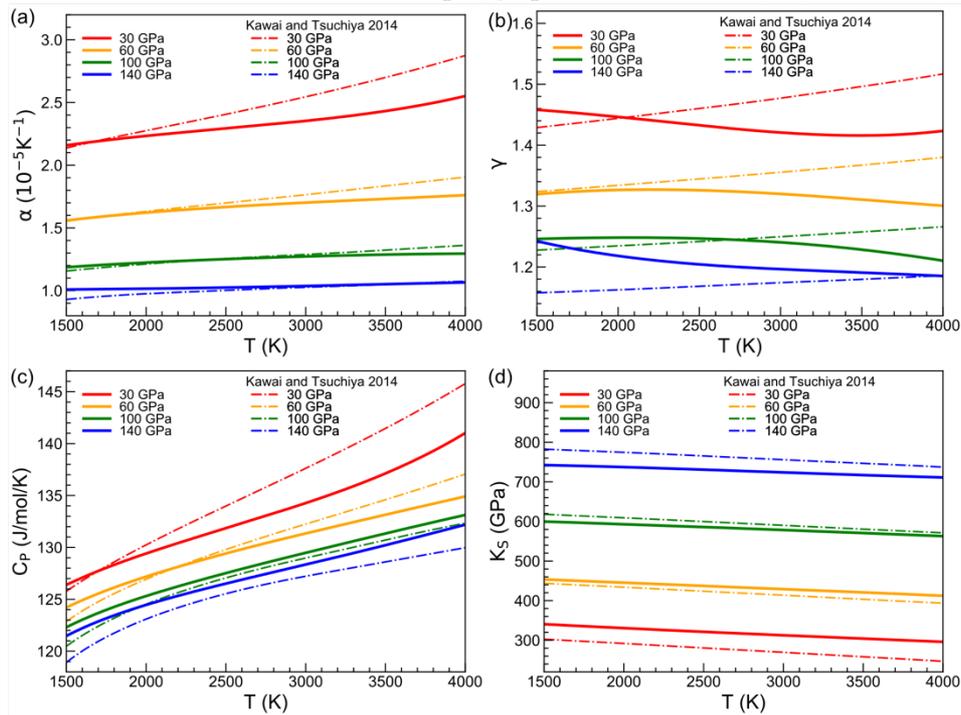

**Figure 5.** (a) Thermal expansivity ($\alpha$), (b) thermodynamic Grüneisen parameter ($\gamma$), (c) isobaric heat capacity ($C_P$), and (d) adiabatic bulk modulus ($K_S$) versus $T$ at a series of $P$ of cubic CaPv, compared with results obtained by the MGD formulation [28]. (Results used with permission from [15], Copyright APS.)

*3.1.3. Metallic Fe.* The method has also been applied to determine the melting properties of hexagonal close-packed (*hcp*) Fe at conditions of the Earth's inner core boundary (330 GPa, 5500 to 6500 K). These properties have great significance in Earth sciences. Yet, precise determination of melting properties is challenging as it requires highly accurate free energies: an error of 10 meV/atom shifts the predicted melting temperature by ~100 K. Accordingly, melting properties are often evaluated via the formally exact TI approach. However, performing TI involves a series of extra AIMD simulations along the integration path, making it computationally very demanding. By contrast, the phonon quasiparticle method requires only one post-processing of the AIMD trajectory and is much more efficient. Figure 6 summarizes the results of the phonon quasiparticle calculations. The mode-projected VDoS exhibit nice Lorentzian line shapes (figure 6(a)), indicating well-defined phonon quasiparticles. These phonon quasiparticles can exhibit either red or blue frequency shifts with respect to their harmonic counterparts. However, the red-shifts dominate (figures 6(b) and 6(c)), leading to an increase of the vibrational entropy, from the harmonic value of 10.15 $k_B$/atom to 10.48 $k_B$/atom. We also find that Fourier interpolation is critical to get the vibrational entropy in the thermodynamic limit: the ionic entropy associated with the 64-atom supercell (without interpolation) is 10.25 $k_B$/atom, 0.23 $k_B$/atom lower than the converged value. As the entropy change at melting is ~1 $k_B$/atom, such differences would cause profound changes (~20%, 1000 K) in the predicted melting temperature. Remarkably, while the computational procedures of the phonon quasiparticle method and TI are quite different, the Gibbs free energies predicted from the two approaches are nearly identical (figure 6(d), within ±10 meV/atom) [12].

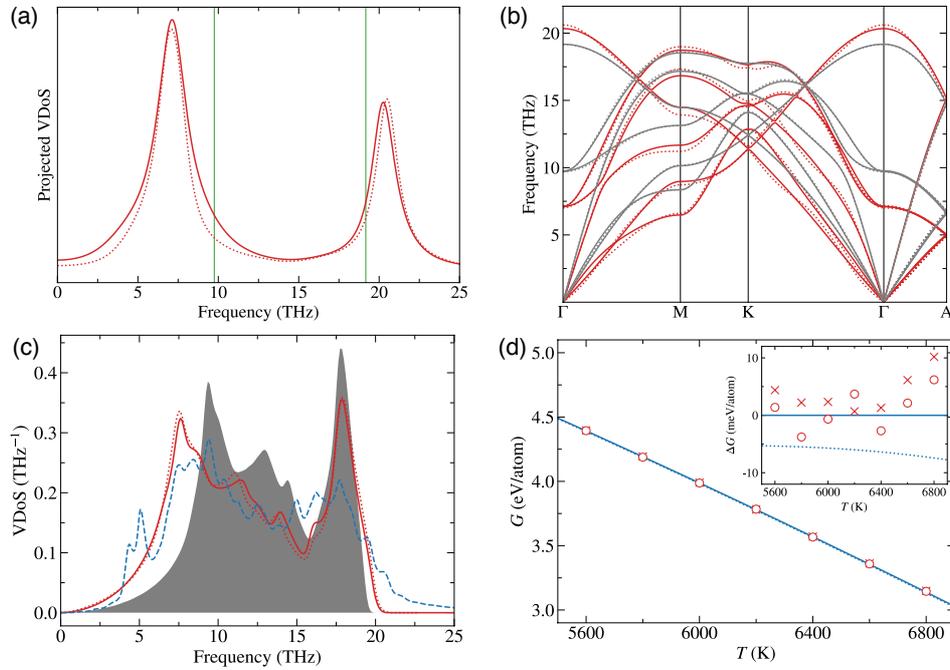

**Figure 6.** Phonon quasiparticles of *hcp* iron at 330 GPa and 6400 K ($V = 6.957$ Å$^3$/atom). (a) Mode projected VDoS at **q** = **Γ** (0, 0, 0). Vertical lines correspond to the frequencies of harmonic TO (left) and LO (right) phonons. (b) Dispersions of phonon quasiparticles (red) and harmonic phonons (grey). (c) Complete VDoS of phonon quasiparticles, where the effective harmonic force constant matrices are from MD involving 64 atoms (4×2×2 supercell) (solid red curve) and 180 atoms (5×3×3 supercell) (red dotted curve). The grey area denotes the VDoS of harmonic phonons, where the harmonic force constant matrix is from the 64-atom supercell. Fourier interpolations are performed on a dense **q**-mesh equivalent to a supercell containing 64000 atoms for both phonon quasiparticles and harmonic phonons. The dashed blue curve denotes the VDoS of MD involving 64 atoms without interpolation. (d) The Gibbs free energy per atom (*G*) in the thermodynamic limit as a function of *T*. Red circles (crosses) denote results from the phonon quasiparticle approach using simulation cells with 64 (180) atoms. Blue solid (dotted) lines denote results from TI using 64 (180) atoms. (Results used with permission from [12], Copyright APS.)

*3.1.4. Strongly anharmonic metallic Be.* Beryllium is an important material with wide applications ranging from aerospace components to x-ray equipment. Yet, a precise understanding of its phase diagram remains elusive. Be assumes an *hcp* structure at relatively low *T*. However, unlike other metals, the stability of the body-centered cubic (*bcc*) phase at high *T* and the associated *hcp/bcc* phase transition are not well understood yet. We have investigated the phase stability of Be using the phonon quasiparticle method that accounts for anharmonic effects. We find that an *hcp* to *bcc* transition occurs near the melting curve at low pressures of $0 < P < 11$ GPa. This also demonstrates the validity of this theoretical framework based on the phonon quasiparticle to study the structural stability and phase transitions in strongly anharmonic metallic materials. Phonon quasiparticles of *bcc* are shown in figures 7(a)–(d). At low *T*, the VAF of the chosen mode does not decay exponentially, and the corresponding power spectrum exhibits a non-Lorentzian line shape, indicating the breakdown of the phonon quasiparticle. At high *T*, e.g., 1000 K and above, the phonon quasiparticle becomes well-defined with exponentially decaying cosine function-shaped VAF and Lorentzian-shaped power spectrum, meaning the *bcc* phase is dynamically stabilized. Accordingly, figure 7(e) displays the anharmonic phonon dispersion at 1000 K free of imaginary frequencies, which is not the case in the harmonic dispersion. Anharmonic phonon dispersion was also obtained for the *hcp* phase and is shown in figure 7(f), where anharmonic effects are nonnegligible. Vibrational entropy and Helmholtz free energies were obtained via equations (7) and (9), and Gibbs free energies were then calculated by $G = F + PV$. By comparing the $G$ of both phases, a pre-melting *hcp* to *bcc* phase transition was discovered [5]. Under high *P-T* conditions, the *hcp/bcc* phase boundary computed with the phonon quasiparticle approach [13] was confirmed by another study using a different approach [31].

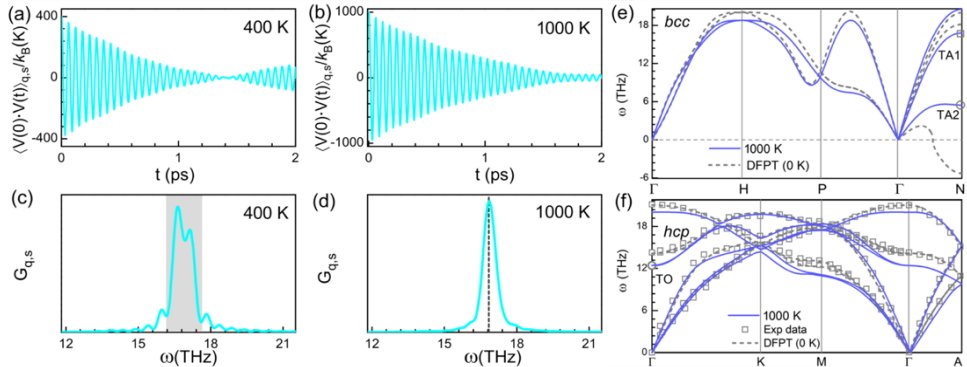

**Figure 7.** Mode-projected VAF of the TA1 acoustic mode at **q** = N with a harmonic frequency of 18.2 THz of *bcc* Be at (a) 400 and (b) 1000 K, respectively. (c) and (d) show the corresponding power spectra. In (c), the shaded area between 16.2 and 17.6 THz covers two major peaks, indicating the breakdown of the phonon quasiparticle picture. In (d), the vertical dashed line at 16.8 THz indicates the frequency of the well-defined phonon quasiparticle. (e) Anharmonic phonon dispersion at 1000 K (solid blue curves) and harmonic phonon dispersion calculated using the DFPT (gray dashed curves) both at $V = 7.81$ Å$^3$/atom, the static *bcc* Be equilibrium volume. The two transverse branches are labeled TA1 and TA2, respectively. (f) Anharmonic phonon dispersion calculated at 1000 K (solid blue curves) and harmonic phonon dispersion calculated using the DFPT (gray dashed curves) both at $V = 7.89$ Å$^3$/atom, the static *hcp* Be equilibrium volume. The experimental data are shown for comparison [32]. (Results used with permission from [5], Copyright APS.)

3.2. Lattice Thermal Conductivity
The phonon quasiparticle method has also been successfully applied to study the lattice thermal conductivity of weakly [18, 33] and strongly [20, 22, 34] anharmonic systems.

*3.2.1. Weakly anharmonic MgSiO₃ perovskite and postperovskite.* At low *T*, the prediction of $\kappa_{lat}$ by equation (10) is usually reliable. However, it fails to recur the high *T* saturation of thermal conductivity, a phenomenon where $\kappa_{lat}$ violates the $1/T$ law as discovered in many materials [35-37]. In order to address the issue, the minimal phonon mean free path (MFP) hypothesis was proposed [6, 36, 38]. It states that the MFP of a phonon, $l_{\mathbf{q}s}$, cannot be shorter than the lattice constants. Otherwise, the phonon picture breaks. With this hypothesis as an amendment to the Peierls-Boltzmann theory, equation (10) yields a minimal lattice thermal conductivity when $l_{\mathbf{q}s}$ approaches the lattice constants at high $T$. As a phenomenology, the minimal phonon MFP has been regarded as valid and useful in various areas. Here we provide a validity check of the minimal MFP hypothesis using the phonon quasiparticle approach and taking MgPv as an example. In our simulation, we have carried out Born-Oppenheimer molecular dynamics (BOMD) [27] on a 2×2×2 supercell of MgPv for a series of $T$ from 300 to 1500 K. The initial atomic structure is first fully relaxed to $P$ = 0 GPa with variable-cell-shape molecular dynamics [39]. The resulting volume is 24.1 cm$^3$/mol.

Figures 8(a) and 8(b) show the obtained mode projected VAF for one optical mode at $\mathbf{q}$ = (0, 0, 1/2) with a harmonic frequency 884 cm$^{-1}$ at $T$ = 300 and 1500 K, respectively. Both display an oscillating behavior. Figures 8(c) and 8(d) show that the corresponding power spectra both display a single peak. From the power spectra, we can also identify the frequency shift $\Delta\omega_{\mathbf{q}s}$ and line width $\Gamma_{\mathbf{q}s}$. At 300 K (1500 K), $\Delta\omega_{\mathbf{q}s} \approx 2$ cm$^{-1}$ (10 cm$^{-1}$) and $\Gamma_{\mathbf{q}s} \approx 7.3$ cm$^{-1}$ (34.5 cm$^{-1}$). Therefore, the criterion for well-defined phonon quasiparticles, $|\Delta\omega_{\mathbf{q}s} - i\Gamma_{\mathbf{q}s}| \ll \omega_{\mathbf{q}s}$, is satisfied both at low $T$ and at high $T$. The good characterization of phonon quasiparticles enables us to investigate the vibrational properties with the concept of phonon. Figure 8(e) shows that the obtained $l_{\mathbf{q}s}$ for the same mode, which broadly follows a $1/T$ law as guided by the solid curve. This is not surprising given a weakly anharmonic system like MgPv. More importantly, beyond certain temperatures, i.e., $T$ = 500, 672, and 680 K, $l_{\mathbf{q}s}$ can be shorter than lattice constants $a_0$, $b_0$, and $c_0$, respectively. Since phonon quasiparticle is valid, i.e., lattice vibrations can be properly depicted in terms of phonon throughout the temperature range considered here, this result suggests that it is unphysical to presume a lower bound in MFP, i.e., the minimal MFP phenomenology does not exist [18].

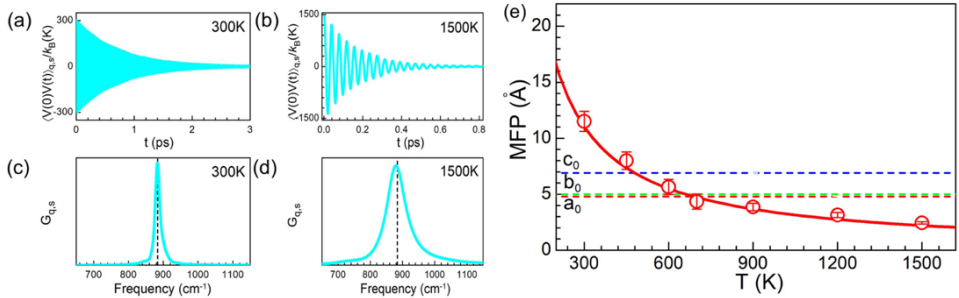

**Figure 8.** The mode-projected VAF (a) at 300 K and (b) 1500 K, respectively, of one optical mode at $\mathbf{q}$ = (0, 0, 1/2). (c) and (d) are the corresponding power spectra. The vertical dashed lines at 884 cm$^{-1}$ indicate the harmonic frequency of this phonon mode. (e) MFP, $l_{\mathbf{q}s}$, of the same mode. Scatters denote the calculated data, and continuous curves having the $1/T$ dependence are the fitted results. The horizontal dashed lines represent the lattice parameters of MgPv, $a_0$, $b_0$, and $c_0$, which define the orthogonal primitive cell with a volume of 24.1 cm$^3$/mol. (Results used with permission from [18], Copyright APS.)

The phonon quasiparticles calculation was also conducted for *Cmcm* MgSiO$_3$ post-perovskite (MgPPv), which is also weakly anharmonic. To overcome the finite-size effects on the lattice thermal conductivity, we relied on the frequency dependence of phonon linewidth [18, 33, 34]. The obtained phonon lifetimes, group velocities (figure 9(a)), and lattice thermal conductivities were compared between the MgPv and MgPPv. $\kappa_{\text{lat}}$ of MgPv as functions of $T$ and $P$ are displayed in figures 9(b) and 9(c), respectively. MgPv's $\kappa_{\text{lat}}$ at 300 K agrees well with a previous theoretical report using perturbative method accounting for up to three-phonon scattering processes [40], which is in line with the weakly anharmonic nature of MgPv. $\kappa_{\text{lat}}$ of MgPPv as functions of $T$ and $P$ are displayed in figures 9(d) and 9(e), respectively. MgPPv's $\kappa_{\text{lat}}$ agrees well with previous experimental measurements at 300 K [41]. Experimental and theoretical discrepancies on MgPv's and MgPPv's $\kappa_{\text{lat}}$ were also discussed [33]. At each temperature, both MgPv's and MgPPv's $\kappa_{\text{lat}}$ vary linearly with pressure. It was also found that MgPPv's $\kappa_{\text{lat}}$ is ~25% higher than that of MgPv. This is attributed to MgPPv's smaller primitive cell

(10 atoms) compared to that of MgPv (20 atoms). Thus, MgPPv has generally larger $v_{\mathbf{q}s}$, as shown in figure 9(a), and higher $\kappa_{\text{lat}}$ [33].

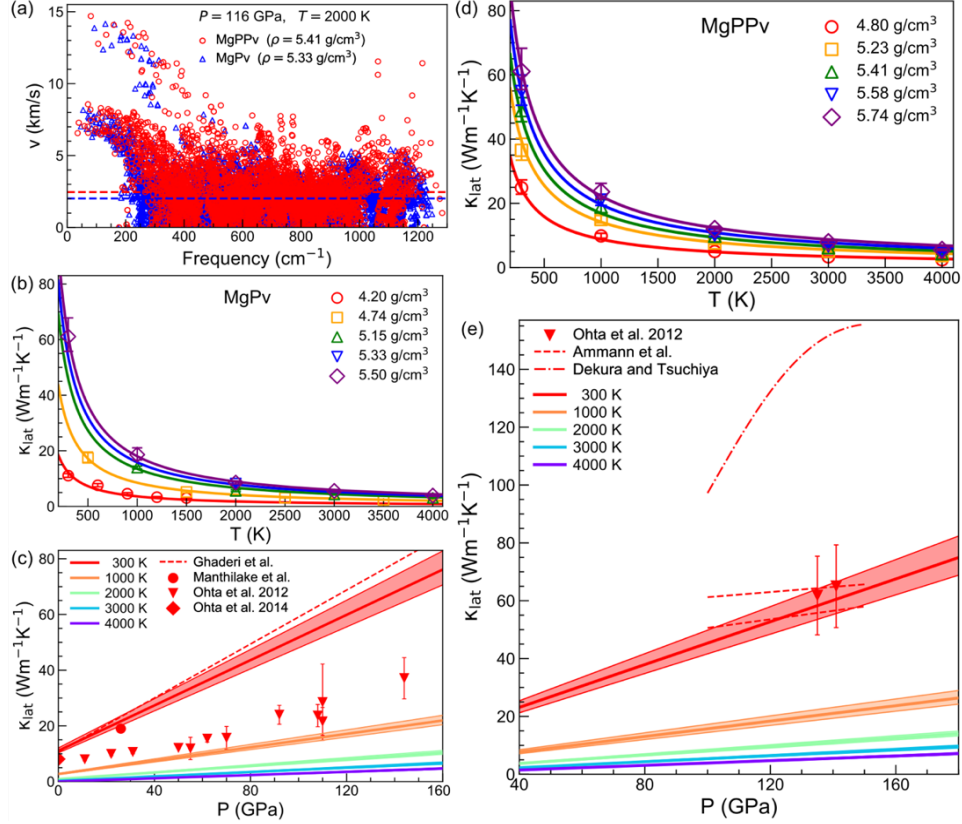

**Figure 9.** (a) $v_{\mathbf{q}s}$ versus $\widetilde{\omega}_{\mathbf{q}s}$ sampled by a 10×10×10 **q**-mesh of MgPPv at $P$ = 116 GPa and $T$ = 2000 K are shown as red circles. $v_{\mathbf{q}s}$ versus $\widetilde{\omega}_{\mathbf{q}s}$ sampled by an 8×8×8 **q**-mesh of MgPv at the same $P$-$T$ conditions are shown as blue triangles for comparison. Dashed lines indicate the average phonon velocities. (b) $\kappa_{\text{lat}}$ versus $T$ of MgPv (symbols) for a series of densities. Error bars show the computational uncertainties. (c) $\kappa_{\text{lat}}$ versus $P$ of MgPv at a series of $T$. Shaded areas indicate the same computational uncertainties as in (b). Symbols with error bars show previous experimental measurements at 300 K [41-43]. The dashed line shows the previous *ab initio* perturbative calculations at 300 K [40]. (d) $\kappa_{\text{lat}}$ versus $T$ of MgPPv (symbols) for a series of densities. Error bars show the computational uncertainties. (e) $\kappa_{\text{lat}}$ versus $P$ of MgPPv at a series of $T$. Shaded areas indicate the same computational uncertainties as in (d). Symbols with error bars show the previous experimental measurements at 300 K [41]. Dashed lines show the previous semiclassical nonequilibrium MD calculations by adopting two different sets of interatomic potentials at 300 K [44]. The dashed-dotted line shows the previous *ab initio* perturbative calculations at 300 K [45]. (Results used with permission from [33], Copyright APS.)

*3.2.2. Strongly anharmonic CaSiO3 perovskite.* CaPv is the last primary lower mantle phase whose thermal conductivity has not been determined, either experimentally or theoretically, due to its strong anharmonicity. The measurements must be carried out at high $P$-$T$ because the cubic phase is unquenchable [4, 46, 47]. Calculations by perturbative methods without renormalization of interatomic force constants are also invalid for cubic CaPv. The static double-well potential deviates from the harmonic one substantially, and harmonic phonons exhibit imaginary frequencies. The phonon quasiparticle approach can overcome such difficulty by extracting well-defined quasiparticle lifetimes at high $T$. Only when phonon quasiparticles exist can equation (10) be used to calculate $\kappa_{\text{lat}}$ within the RTA of the LBTE [34]. Figure 10(a) shows the obtained MFP. The sub-minimal MFP was also observed in such a strongly anharmonic system. Figures 10(b) and 10(c) report the obtained $\kappa_{lat}$ of cubic CaPv as functions of *T* and *P*, respectively. The linear dependence of $\kappa_{lat}$ on pressure was once again observed. Our theoretical predictions for $\kappa_{lat}$ were further confirmed by the high *P-T* thermoreflectance measurements [34, 48, 49].

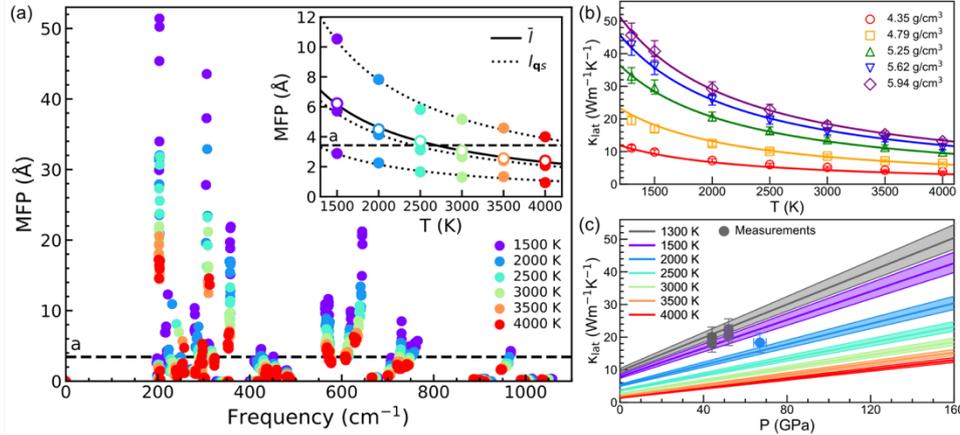

**Figure 10.** (a) MFP, $l_{\mathbf{q}s}$, from phonon quasiparticles sampled by MD simulations using a 3×3×3 supercell (135 atoms) with $\rho$ = 4.79 g/cm³ of CaPv at a series of *T*. The horizontal dashed black lines in both the main and the inserted figures represent the primitive cell lattice parameter, *a*. Insert: Three $l_{\mathbf{q}s}$ (solid circles) versus *T* are showcased. One remains longer than *a* (top), one remains shorter than *a* (bottom), and the other crosses the lattice parameter (middle) in the *T* range investigated. Average MFP, $\bar{l}$ (hollow circles), is also found to cross *a* as *T* increases. We find no lower-bound limit on phonon MFP in CaPv. The dotted and solid black curves show the *1/T* fits [18]. (b) $\kappa_{lat}$ versus *T* of CaPv (symbols) at several densities. Error bars show the computational uncertainties. (c) $\kappa_{lat}$ versus *P* of CaPv at a series of *T*. Shaded areas indicate the same computational uncertainties as in (b). Solid symbols with error bars are experimental data by the high *P-T* thermoreflectance measurements [34, 48, 49]. (Results used with permission from [34], Copyright APS.)

## 4. Conclusion

The phonon quasiparticle approach addresses lattice anharmonicity by extracting renormalized frequencies and phonon lifetimes from AIMD simulations without explicitly expressing the higher-order interatomic force constants. In principle, it captures full anharmonicity and can overcome finite-size effects typical of AIMD simulations by offering complete phonon quasiparticle dispersions through Fourier interpolation. It has been demonstrated [4, 5, 10, 12, 13, 15, 18-22, 33, 34] to be valid and powerful in studying thermodynamics and lattice thermal conductivities of both weakly and strongly anharmonic materials when phonon quasiparticles are well defined.

**References**
[1]   Souvatzis P, Eriksson O, Katsnelson M I and Rudin S P 2008 *Phys. Rev. Lett.* **100** 095901
[2]   Errea I, Rousseau B and Bergara A 2011 *Phys. Rev. Lett.* **106** 165501
[3]   Hellman O, Abrikosov I A and Simak S I 2011 *Phys. Rev. B* **84** 180301(R)
[4]   Sun T, Zhang D-B and Wentzcovitch R M 2014 *Phys. Rev. B* **89** 094109
[5]   Lu Y, Sun T, Zhang P, Zhang P, Zhang D-B and Wentzcovitch R M 2017 *Phys. Rev. Lett.* **118** 145702
[6]   Kittel C 1949 *Phys. Rev.* **75** 972
[7]   Wallace D C 1965 *Phys. Rev.* **139** A877
[8]   Carrier P, Wentzcovitch R M and Tsuchiya J 2007 *Phys. Rev. B* **76** 064116
[9]   Klemens P G 1958 *Solid State Phys.* **7** 1
[10]  Zhang D-B, Sun T and Wentzcovitch R M 2014 *Phys. Rev. Lett.* **112** 058501
[11]  Sun T, Shen X and Allen P B 2010 *Phys. Rev. B* **82** 224304

[12]  Sun T, Brodholt J P, Li Y and Vočadlo L 2018 *Phys. Rev. B* **98** 224301
[13]  Xian J-W, Yan J, Liu H-F, Sun T, Zhang G-M, Gao X-Y and Song H-F 2019 *Phys. Rev. B* **99** 064102
[14]  Mermin N D 1965 *Phys. Rev.* **137** A1441
[15]  Zhang Z and Wentzcovitch R M 2021 *Phys. Rev. B* **103** 104108
[16]  Ziman J M 2001 *Electrons and Phonons: The Theory of Transport Phenomena in Solids.* (Oxford: Oxford Univ. Press)
[17]  Sun T and Allen P B 2010 *Phys. Rev. B* **82** 224305
[18]  Zhang D-B, Allen P B, Sun T and Wentzcovitch R M 2017 *Phys. Rev. B* **96** 100302(R)
[19]  Zhang Z, Zhang D-B, Sun T and Wentzcovitch R M 2019 *Comput. Phys. Commun.* **243** 110
[20]  Lu Y, Sun T and Zhang D-B 2018 *Phys. Rev. B* **97** 174304
[21]  Lu Y, Zheng F, Yang Y, Zhang P and Zhang D-B 2019 *Phys. Rev. B* **100** 054304
[22]  Lu Y, Zheng F, Yang Y, Zhang P and Zhang D-B 2021 *Phys. Rev. B* **103** 014304
[23]  Gillet P, Guyot F and Wang Y 1996 *Geophys. Res. Lett.* **23** 3043
[24]  Gillet P, Daniel I, Guyot F, Matas J and Chervin J C 2000 *Phys. Earth Planet. Inter.* **117** 361
[25]  Durben D J and Wolf G H 1992 *Am. Mineral.* **77** 890
[26]  Baroni S, Gironcoli S d, Corso A D and Giannozzi P 2001 *Rev. Mod. Phys.* **73** 515
[27]  Wentzcovitch R M and Martins J L 1991 *Solid State Commun.* **78** 831
[28]  Kawai K and Tsuchiya T 2014 *J. Geophys. Res. Solid Earth* **119** 2801
[29]  Jackson I and Rigden S M 1996 *Phys. Earth Planet. Inter.* **96** 85
[30]  Anderson O L 1995 *Equations of State of Solids for Geophysics and Ceramic Science.* (New York: Oxford University Press)
[31]  Wu J, González-Cataldo F and Militzer B 2021 *Phys. Rev. B* **104** 014103
[32]  Stedman R, Amilius Z, Pauli R and Sundin O 1976 *J. Phys. F* **6** 157
[33]  Zhang Z and Wentzcovitch R M 2021 *Phys. Rev. B* **103** 144103
[34]  Zhang Z, Zhang D-B, Onga K, Hasegawa A, Ohta K, Hirose K and Wentzcovitch R M 2021 *Phys. Rev. B* **104** 184101
[35]  Auerbach A and Allen P B 1984 *Phys. Rev. B* **29** 2884
[36]  Cahill D G, Braun P V, Chen G, Clarke D R, Fan S, Goodson K E, Keblinski P, King W P, Mahan G D, Majumdar A, Maris H J, Phillpot S R, Pop E and Shi L 2014 *Appl. Phys. Rev.* **1** 011305
[37]  Roufosse M C and Klemens P G 1972 *J. Geophys. Res.* **79** 703
[38]  Grimvall G 1999 *Thermophysical Properties of Materials.* (Amsterdam: North-Holland)
[39]  Wentzcovitch R M 1991 *Phys. Rev. B* **44** 2358
[40]  Ghaderi N, Zhang D-B, Zhang H, Xian J, Wentzcovitch R M and Sun T 2017 *Sci. Rep.* **7** 5417
[41]  Ohta K, Yagi T, Taketoshi N, Hirose K, Komabayashi T, Baba T, Ohishi Y and Hernlund J 2012 *Earth Planet. Sci. Lett.* **349–350** 109
[42]  Manthilake G M, Koker N d, Frost D J and McCammon C A 2011 *Proc. Natl. Acad. Sci. U.S.A.* **108** 17901
[43]  Ohta K, Yagi T and Hirose K 2014 *Am. Mineral.* **99** 94
[44]  Ammann M W, Walker A M, Stackhouse S, Wookey J, Forte A M, Brodholt J P and Dobson D P 2014 *Earth Planet. Sci. Lett.* **390** 175
[45]  Dekura H and Tsuchiya T 2019 *Geophys. Res. Lett.* **46** 12919
[46]  Komabayashi T, Hirose K, Sata N, Ohishi Y and Dubrovinsky L S 2007 *Earth Planet. Sci. Lett.* **260** 564
[47]  Thomson A R, Crichton W A, Brodholt J P, Wood I G, Siersch N C, Muir J M R, Dobson D P and Hunt S A 2019 *Nature* **572** 643
[48]  Hasegawa A, Yagi T and Ohta K 2019 *Rev. Sci. Instrum.* **90** 074901
[49]  Okuda Y, Ohta K, Hasegawa A, Yagi T, Hirose K, Kawaguchi S I and Ohishi Y 2020 *Earth Planet. Sci. Lett.* **547** 116466